# Subtraction of "accidentals" and the validity of Bell tests

Caroline H Thompson*
Department of Computer Science, University of Wales, Aberystwyth,
SY23 3DB, U.K.
(October 24, 2018)

In some key Bell experiments, including two of the well-known ones by Alain Aspect, 1981-2, it is only after the subtraction of "accidentals" from the coincidence counts that we get violations of Bell tests. The data adjustment, producing increases of up to 60% in the test statistics, has never been adequately justified. A straightforward realist model, assuming pulsed classical light and giving good fit to the unadjusted data, is discussed. In the light of this, and of the other known Bell test "loopholes", the claim that the universe is fundamentally nonlocal needs re-assessment.

03.65.Bz, 03.65.Sq, 03.67.*, 42.50.Ct

In early "EPR" experiments, testing ideas inspired by Einstein, Podolsky and Rosen as set out in their famous 1935 paper [1], it is clear there were doubts regarding the validity of Bell tests using data that had been "adjusted for accidentals". In a paper by Freedman and Clauser [2], for instance, the Bell test was conducted using both raw and adjusted data. In this particular case, the difference was unimportant. The emission rates used were exceptionally low, and so (as will be shown later) it is not surprising that accidentals were very low, only about 1 in 40 of the detected coincidences.

But, in 1981-2 [3], Aspect, possibly following earlier precedents, subtracted accidentals in experiments in which they formed around 25% of his coincidences. The procedure was queried in 1983 by Marshall, Santos and Selleri [4], and later by Wesley [5]. Aspect and Grangier responded to the 1983 challenge [6]. They employed theoretical arguments, though, that involved many assumptions [7], and supported these by a small amount of additional data. This was from Aspect's "two-channel" ('+' and '−' from each polariser) experiment only. For this experiment, the raw data does indeed produce Bell violations, but the Bell test used here (using $S_{Std}$, Appendix A) is one that is readily violated if the "fair sampling" assumption, for example, is false [8]. His other, "single channel", experiments used a different – and, in my view, generally superior – Bell test (using $S_C$: see Appendices A and B, which include a very simple derivation). So far as can be judged from analysis of data from his PhD thesis [9], the raw data for these does *not* violate the test.

The magnitude of the problem has never been publicised. Somehow it became customary to adjust the data, and to publish no explanation. Wesley had to use considerable imagination to deduce the order of magnitude of the effect on Aspect's Bell tests, as he had access only to information from the Physical Review Letters papers [3]. From about 1983 till 1998, some experimenters (for example, Rarity and Tapster in 1994 [10] and Tittel et al in 1997 [11]) adjusted data as a matter of routine, whilst others [12] did not.

Data on accidental rates is not always given, but Tittel's 1997 paper is an exception, including a graph from which it is easy to see that the rate is about 30%. The adjustment changes the visibility ($S_V$, Appendix A) from about 45% to 82%, an increase of about 60% and sufficient to bring the value above the Bell limit of 71%! This is the paper that revived my interest in the subject, as I had hitherto assumed that the practice would have been abandoned with the switch to "PDC" (Parametric Down-Conversion) sources in place of atomic cascades. (Aspect's justification, involving the assumed independence of emission events, does not so readily apply. See later.) I corresponded with Tittel et al, and placed a paper in the Los Alamos quantum physics archive [13]. Many, if not all, experimenters in the field now recognise that it is the raw data that *must* be used in Bell tests [14]. The adjustment alters the test statistic. The amount varies, but simple algebra shows that the direction is always in favour of the quantum theory prediction.

## I. EPR EXPERIMENTS: THEORY AND PRACTICE

Now, as is well documented in the literature [15–17], in an idealised "EPR experiment" a source emits pairs ($A$ and $B$) of correlated quantum particles. They are sent to analysers (polariser/photomultiplier combinations, for example) whose settings can be chosen by the experimenter. If an analyser detects its particle, it clocks up a 1. If both do this simultaneously, a coincidence is scored. By repeating this experiment using several different settings, the manner in which coincidence rates vary as we vary the settings can be studied, and data extracted for the conduct of Bell tests. The latter were devised so as to distinguish the quantum theory (QT) prediction from models that rely on purely local effects. QT predicts peculiarly high correlations between the *A* and *B* detections, ones that cannot be explained without assuming "non-local entanglement". "Realist" explanations, relying only on the effects of "common causes", predict (for *perfect* experiments!) lower correlations.

In practice, almost all EPR experiments to date have used light. The quantum theory interpretations have assumed that the light emitted, whether by "radiative atomic cascades" or by PDC, is produced in pairs of "photons". But the two detections are not registered at quite the same moment, either because they were not emitted quite simultaneously or because of the imperfect time-resolution of the detectors. (In practice, the fact that in PDC the two photons are in theory emitted at the same time makes only a slight quantitative difference [18].) It is not, therefore, entirely obvious which detections represent correlated pairs. The problem is exacerbated by the fact that the detectors are far from perfect, with "efficiency" in the early experiments – and even some recent ones – of 5% or lower. In order to estimate the number of paired detections, the times between successive *A* and *B* counts are analysed to give a time spectrum (see example, Fig. (1)). A "coincidence window" is defined by the experimenter (for the illustrated example it was from 3 ns before the peak to 17 ns after it) and all results falling between these limits are taken to be coincidences.

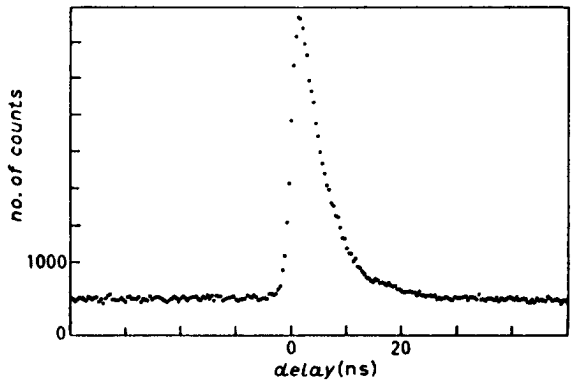

FIG. 1. Time spectrum from Aspect's thesis. (Actual runs would have had considerably greater scatter as they were over shorter periods.)

Thus this integral from −3 ns to +17 ns gives us our raw coincidence counts. Clearly, in the ideal situation, with pairs produced at large intervals, we would expect the time spectrum to be zero at all points outside our time window. It is not. There is a strong temptation to assume that the flat regions to either side of the peak represent some kind of steady "accidental" rate, and that this should be subtracted from our integral to give "true" coincidences. But is this correct?

## II. ESTIMATION AND LOGIC OF "ACCIDENTALS"

There are two main methods of estimating so-called "accidentals". The most common is to take the stream of electronic data generated by the detector on one side and delay it by, say, 100 ns. This is sufficient to destroy any synchronisation, and so coincidences measured now must necessarily represent chance, or accidental, ones [19]. The other method, reported to give almost identical results, is to take the two "singles rates" and multiply them. After adjusting for window size, this also gives a natural measure of accidental rate. The objectivity of the procedure is not here in question, only its relevance – whether or not it gives a fair estimate for the region near "zero", as well as for points well away from it.

For it is not logical to consider the region near the "zero" time-difference point as the same as all others. It represents the simultaneous detection of two light signals that were emitted together or, possibly, genuine signals matched with "noise". There are real possibilities that are being ignored if the accidental rate is assumed the same at zero as elsewhere.

- The source may not in fact be capable of emitting pairs of signals without a short break in between. Ignoring noise, therefore, perhaps all detections within the coincidence window *must* come from the same emission event. The latter will, in a classical theory, form a prolonged pulse, so that detections are spread over time only because different parts of the pulse are detected at random [20].

- Probabilities of detection may not be additive. Noise added to a strong signal may have little effect, whilst added to a weak one it causes detection.

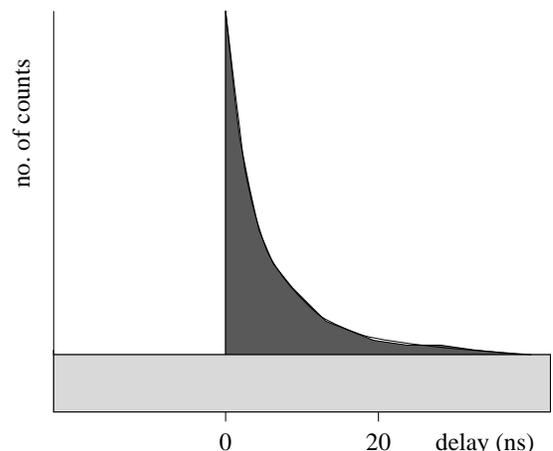

FIG. 2. Model of time spectrum, from Aspect's thesis. Dark shading: "true"; Light shading: "accidental" coincidences.

Aspect assumed [9] that every one of the thousands of atoms in his source region acted independently, so that emissions from different atoms could occur arbitrarily close in time, making "wrong" pairings (almost) as likely at zero as at any other time difference. He devoted several pages of his thesis to the matter, illustrating his model as shown in Fig. (2). His theory left him free to subtract "accidentals", and hence free to ignore them in his choice of experimental parameters. He chose to follow Stuart Freedman [21], minimising a "quality factor" that amounted to using the minimum running time for the experiment to achieve statistical significance for his results. This criterion led him to chose a fairly high emission rate, with consequent high accidental rates, as these are proportional to the product of the rates in the two streams.

It is of interest to note that Aspect would have expected to be able to produce the same Bell violations with no need for adjustment for accidentals if he had used a reduced emission rate. This is because, though the accidentals vary as the product of rates, $N^2$, the coincidence rate should vary directly with $N$ itself. Indeed, as he states in his thesis and in a footnote in [6], this relationship has been confirmed experimentally. As mentioned earlier, Freedman in 1972 had a low emission rate (or, at least, a low detection rate) and hence sufficiently low accidentals for them not to affect his conclusion. It follows that a much more satisfactory way of countering challenges as to the validity of the adjustment would have been to conduct experiments at lower emission rates, rather then rely on theory.

Tittel [11] assumed that the majority of the accidentals in his long-distance Bell tests were due to noise accumulated in transit, along the several kilometers of fibre-optic cable. He states that a "dark count" of 100 kHz was included in a singles rate of 170 kHz. (There is a confusion of terminology here: he means, presumably, the count obtained when the signal is cut off at source but noise from the environment of the cable is not excluded. It is sometimes necessary to distinguish this noise from the basic dark count registered by a photomultiplier in the absence of any input at all.) For a PDC source, both QT and classical theory lead to the assumption that only one pair of signals is produced at a time. Aspect's argument does not so readily apply, as the chance of two emissions occurring within the time window so that accidental wrong pairing can take place is assumed negligible.

## III. QUANTUM THEORY AND LOCAL REALIST PREDICTIONS

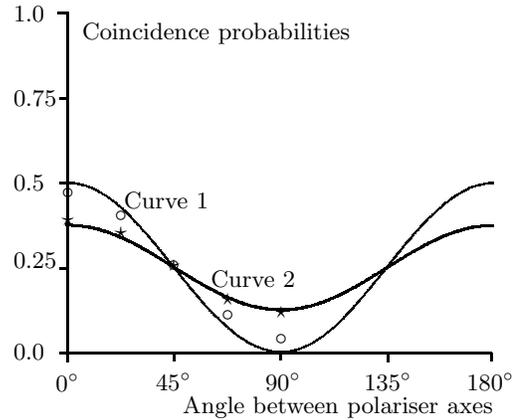

FIG. 3. Principal predicted "coincidence curves" for the ideal case, and actual data from Aspect's 1981 experiment. Curve (1): quantum theory; curve (2): the basic local realist prediction. ⋆ indicates raw data; ∘ adjusted.

The quantum theory prediction for the probability of (same-channel) coincidences for a perfect experiment, using plane polarised light, is well known to be

$$P_{AB,Q} = \frac{1}{2} \cos^2 \phi, \qquad (1)$$

curve (1) of Fig. 3, where $\phi$ is the angle between the polariser axes.

The standard local realist prediction when using polarised light is not in fact so very different (curve (2)). The plotted data points will be discussed in the next section.

Note that for the purposes of actual Bell tests, it is the value of the minimum that is crucial. For light, indeed, a formal Bell test is not needed to discriminate between quantum theory and this model. A simple test of visibility (see Appendix A), using just the maximum and minimum values, will suffice. Whereas for the idealised Stern-Gerlach experiment quantum theory and local realist models agree as to the maximum and minimum, both saying that the coincidence probability for a given channel is 0.5 for parallel detectors, 0 for orthogonal ones, for light the realist model gives a lower maximum and non-zero minimum. That it *must* be non-zero *in this perfect case* follows directly from the assumptions, as explained later.

The basic assumptions for a perfect experiment are that there should be no preferred polarisation direction (the source is "rotationally invariant") and, for the quantum theory version, that "quantum efficiency" of the detectors is 1. It is also taken as read that there are no accidentals or other ambiguities – pairs are clearly identified. It is further assumed that we have symmetry between the two sides. The absence of preferred direction then implies that the coincidence curve will be a function of the angle between the polarisation axes of the two detectors.

### A. Quantum theory prediction

For a full treatment of the QT prediction, the reader is referred to texts such as that by Mandel and Wolf [22]. My own understanding of the general idea is as follows. A joint wave function for the two "photons" is established, taking account of their "entanglement", then the "projection postulate" is employed to translate this into the appropriate form to allow for the two polariser settings.

There is some conceptual difficulty, as it is not possible to assign a specific polarisation direction to a photon until it is detected. Until detection, it exists in a superposition of vertically and horizontally polarised states, the definition of "vertical" being at the discretion of the observer. The theory is tied to the idea that we are dealing with dichotomous observables, and instruments that produce outcomes that are either "+" or "−", never "some of each" or "nothing". The joint state is known as the "singlet state", represented in symbols by formulae such as:

$$|\psi\rangle = \frac{1}{\sqrt{2}}(|\updownarrow\rangle|\updownarrow\rangle + |\leftrightarrow\rangle|\leftrightarrow\rangle). \qquad (2)$$

The projection postulate enables one to transform this wave function into predicted results for polarisers set with axes at relative angle $\phi$. The formula *can* be adapted slightly to allow for imperfect instruments – the occasional absorption of a photon by a polariser, say – but the adaptation requires specialist skill. All the variants I have encountered (for example, in Aspect's work and papers such as that by Lepore and Selleri [23]) appear to be constrained to depend on $\phi$ only through $\cos^2$ terms, never through more general functions.

### B. Realist prediction

The realist prediction that I am considering follows from the assumption that the two emitted signals are not photons but short pulses of classical light. They have a common polarisation direction, which forms the "hidden" or "common cause" variable that causes the correlation that gives rise to the coincidence pattern of curve (2). For the perfect case, the assumptions made are that:

- the intensity of the light is reduced by the polarisers following Malus' Law [24], the intensity for light polarised parallel to the axis being unaffected by the polariser.

- the detectors have a linear response to (electromagnetic) intensity, i.e. if intensity is $I$, the probability of a count is proportional to $I$. Light that does not pass through a polariser or that is polarised parallel to its axis is detected with probability 1.

- the "factorability" assumption: for every fixed polarisation direction $\lambda$, the probability of coincidence is the product of the probabilities of detection for the two signals separately.

Interpreting the above, we have, for polarisation axes at angles $a$ and $b$ respectively:

$$\begin{aligned} p(a,\lambda) &= \cos^2(\lambda - a) \\ p(b,\lambda) &= \cos^2(\lambda - b), \end{aligned} \qquad (3)$$

for the "singles" probabilities, giving probability of a coincidence for the polarisation direction $\lambda$

$$p_{AB,R}(a,b,\lambda) = p(a,\lambda)p(b,\lambda). \qquad (4)$$

Integrating over all polarisation directions, this yields average probability of a coincidence

$$P_{AB,R}(a,b) = \int_0^\pi \frac{d\lambda}{\pi} \cos^2(\lambda - a) \cos^2(\lambda - b) \qquad (5)$$

This expression can be integrated using high school maths (I show details in Appendix C). The listed assumptions, though, can readily be relaxed to allow for experimental imperfections without sacrificing the principles of locality. Thus there is a wide class of variants of the above, in which the weighting factor $\frac{1}{\pi}$ is replaced by a function depending on $\lambda$, or the $\cos^2$ terms replaced by different functions. These are likely to require numerical integration. Even the last, factorability, assumption, often taken to be the definition of locality, can be relaxed slightly as it is not in practice always realistic. It will not hold exactly if there are synchronisation problems [13,25,26], and in this case full computer simulation may be needed. Some of the results will be indistinguishable experimentally from curve (1). But this is by the way (it is partly covered by papers such as Marshall, Santos and Selleri's 1983 one [4] and more recent contributions by Gilbert and Sulcs [27] and Vladimir Nuri [28]). The purpose of the current paper is to discuss the matter of "accidentals".

It is of fundamental importance to realise that the main difference between the QT and realist predictions *for the ideal case* is the non-zero minimum of the realist one. If we assume both adherence to Malus' Law and a uniform distribution of polarisation directions, the fact that *all* $\lambda$ values (other than exactly 90° to the axis) are detected to some degree means the minimum of the average cannot be zero. To see this clearly, consider the case of polarisers set orthogonally, for which the minimum is achieved. In our realist model, take the subensemble of emissions polarised at $\lambda = 45°$ to both axes. The individual probabilities of detection in the + channel are both 1/2, and the coincidence probability 1/4. This is not zero, and nor are any of the other contributions for the other values of $\lambda$, apart from a set of measure zero – that for $\lambda$ *exactly* 0° or 90°. As the integration involves no negative contributions, the result must be greater than zero (in fact, 1/8).

## IV. THE EXPERIMENTAL CONSEQUENCES OF SUBTRACTION OF ACCIDENTALS

Let us now consider some data from real experiments.

### A. Aspect's 1981 experiment

Aspect's first experiment, a single-channel one reported in 1981, is the only one for which I have full data on accidentals. The data in table I is a summary of that presented in his thesis in raw form.

| $\phi$ | 0 | 22.5° | 45° | 67.5° | 90° | $z$ | $Z$ | $S_V$ | $S_C$ |
|---|---|---|---|---|---|---|---|---|---|
| Raw coincidences | 96 | 87 | 63 | 38 | 28 | 126 | 248 | 0.55 | -0.121 |
| "Accidentals" | 23 | 23 | 23 | 23 | 23 | 46 | 90 | | |
| Adjusted values | 73 | 64 | 40 | 16 | 5 | 81 | 158 | 0.88 | 0.096 |

TABLE I. Effect of standard adjustment for "accidental" coincidences. $z$ is rate with one polariser absent, $Z$ that with both absent. $S_V$ = visibility. See Appendix for definition of $S_C$. Under local realism, it cannot exceed zero. (Derived from table VII-A-1 of Aspect's thesis, relating to his 1981 single-channel experiment.)

The pattern of accidentals found here is as expected. If that for both polarisers present is $A$, then, because removal of a polariser doubles the beam intensity (at least approximately), the coincidence rate with one removed is expected to be $2A$, and that with both removed $4A$. The effect of subtraction of accidentals is to shift all points downwards, reducing the minimum to nearly zero. After "normalisation" by division by the value with both polarisers removed, the *adjusted* data is in reasonable agreement with the quantum theory prediction (see Fig. 3). The same normalisation brings the *raw* data, however, into very good agreement with the realist prediction. Both predictions can be improved by allowing for experimental imperfections.

### B. Tittel et al: Quantum Correlations over 10 k, Geneva, 1997

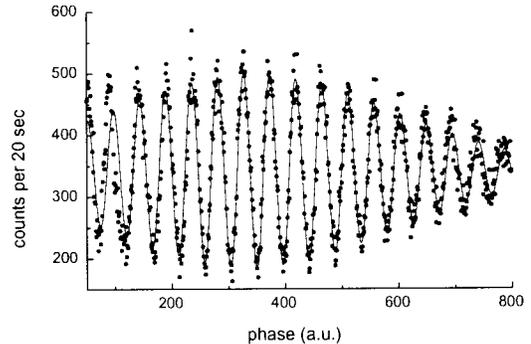

FIG. 4. Coincidence rates against phase shift, from Tittel's 1997 10 k experiment in Geneva.

The data from Tittel et al's 1997 experiment [11] is aptly summarised in the graph of Fig. (4). Note that the vertical axis starts at 150, the "accidental" rate that is subtracted before analysis. The graph represents a coincidence curve taken over many periods, modulated by a function related to the coherence length of the "photons". It was assumed that accidentals were due mainly to noise introduced along the 10 k long optical fibres linking source to detectors. It is clear from the graph that the adjustment was large. I estimate that it would have changed the visibility from about 0.5 to the value of 0.816 that was reported.

## V. DISCUSSION

Some of the key analyses presented for public view have not followed the accepted principle that sufficient information should be given for the reader to make an unbiased assessment. Assumptions have been made that, had they been clearly stated, would have been much more widely challenged. Independence of emission events, for example, is a quantum theory assumption of long standing, yet has it ever been proved? In experiments of fundamental significance, where the very possibility of providing rational explanations for quantum phenomena is at stake, it is not acceptable.

This all seems now to be generally agreed. A reassessment is evidently required, looking not only at the effect of data adjustment but conducting further comprehensive experiments to investigate the effects of altering parameters such as emission rates, beam attenuation, detector properties and coincidence windows. The importance of the known but much-neglected "loopholes" would then, I believe, become apparent. It appears likely that such investigation would result in a very considerable reduction in claims to have observed non-local phenomena. Indeed, they might be eliminated entirely, allowing a return to the view that the world is, after all, obeying local causal rules, even at the quantum level.

## APPENDIX A: BELL TESTS

I give in table II formulae for the Bell tests that are commonly found in actual (optical) experiments. More general formulae are needed if the source lacks rotational invariance or there are asymmetries. (It is no accident that they bear little resemblance to the inequalities originally invented by Bell, as the latter were specifically designed to deal with "spin-1/2" particles, using Stern-Gerlach magnets, and depended on the assumption that all particles were detected.)

| | Test Statistic | Upper Limit | Auxiliary Assumption |
|---|---|---|---|
| Standard | $S_{Std} = 4(\frac{x-y}{x+y})$ | 2 | Fair sampling |
| Visibility | $S_V = \frac{max-min}{max+min}$ | 0.71 | " |
| CHSH | $S_C = 3\frac{x}{Z} - \frac{y}{Z} - 2\frac{z}{Z}$ | 0 | No enhancement |
| Freedman | $S_F = \frac{x-y}{Z}$ | 0.25 | " |

TABLE II. Various Bell inequalities, for rotationally invariant, symmetrical, factorisable experiments. $x = R(\pi/8)$, $y = R(3\pi/8)$, $z = R(a, \infty)$ and $Z = R(\infty, \infty)$, using the usual terminology in which $R$ is coincidence rate (for ++ coincidences, in the two-channel case), $a$ is polariser setting, and $\infty$ stands for absence of polariser.

The first two tests are, I believe, always biased [8,13]. For a valid Bell test, the denominator should be the number of pairs emitted, and the values used are very much too small. The last two tests may be less biased, using the coincidence counts with no polarisers present as denominator. They do suffer, though, from the need to assume "no enhancement" (the presence of a polariser never, for any hidden variable value, *increases* the chance of detection). This is considered by some [4] to be a serious drawback, but my personal view is that it is likely to be of very much less importance than the "fair sampling" failures that can so easily bias other tests. The derivation of the CHSH test is very straightforward and instructive. As it is rarely reproduced, I give it in the next section.

## APPENDIX B: DERIVATION OF THE (SINGLE-CHANNEL) CHSH INEQUALITIES

Clauser and Horne, in their excellent paper of 1974 [29], give elegant derivations of two inequalities. Both are important. The first makes probably the least possible assumptions, but is of use only for near-perfect detectors. The second (using $S_C$) is the one that is used in practice, for example in Freedman's and the first and last of Aspect's experiments [2,3]. It is valid under a wide range of assumptions, even when detector efficiencies are low. Various ideas in this 1974 paper represent notable advances over their 1969 one with Shimony and Holt [30] and it is unfortunate that the latter seems much more widely known.

In hindsight, it would appear that the treatment perhaps does not emphasise sufficiently the possibilities for assumption failures. Two points in particular come to mind. Firstly, they take for granted "factorability", and this may fail. One of the footnotes of their 1974 paper in fact guards against this, though they do not state it explicitly: their footnote [9] gives conditions on coincidence window sizes and pair separation times that should ensure that pairs can be identified unambiguously (and, incidentally, that accidentals are negligible), and if these are satisfied factorability should follow. Secondly, they imply that "rotational invariance" can satisfactorily be ascertained by experiment. For a rigorous confirmation of invariance, however, a considerable amount of extra experimentation would be needed, so that it would be quicker and simpler to make no such assumption and to use the full version of the test. As Bell tests are quite sensitive to invariance failures, and as QT and realist models of a given setup may not agree on this point, it should not be assumed lightly.

To return to Clauser and Horne's exposition: I shall quote in full the derivation of the first test, then briefly outline the second, which follows a similar method.

Starting at page 528 we find the following (in their own words apart from equation numbers):

## 1. Derivation of the "Minimum-assumption CHSH" inequality

... in this section, we derive a consequence of [the factorability assumption] which is experimentally testable without $N$ being known, and which contradicts the quantum-mechanical predictions.

Let $a$ and $a'$ be two orientations of analyzer 1, and let $b$ and $b'$ be two orientations of analyzer 2. The inequalities

$$0 \leq p_1(\lambda, a) \leq 1,$$
$$0 \leq p_1(\lambda, a') \leq 1,$$
$$0 \leq p_2(\lambda, b) \leq 1, \quad \text{(B1)}$$
$$0 \leq p_2(\lambda, b') \leq 1$$

hold if the probabilities are sensible. These inequalities and the theorem [see below (Appendix A of original paper)] give

$$-1 \leq p_1(\lambda, a)p_2(\lambda, b) - p_1(\lambda, a)p_2(\lambda, b')$$
$$+ p_1(\lambda, a')p_2(\lambda, b) + p_1(\lambda, a')p_2(\lambda, b')$$
$$- p_1(\lambda, a') - p_2(\lambda, b) \leq 0 \quad \text{(B2)}$$

for each $\lambda$. Multiplication by $\rho(\lambda)$ and integration over $\lambda$ gives [assuming factorability]

$$-1 \leq p_{12}(a,b) - p_{12}(a,b') + p_{12}(a',b) + p_{12}(a',b')$$
$$- p_1(a') - p_2(b) \leq 0 \quad \text{(B3)}$$

as a necessary constraint on the statistical predictions of any OLT [Objective Local Theory]. If, for some reason such as rotational invariance, it is found experimentally that $p_1(a)$ and $p_2(b)$ are constant, and that $p_{12}(a,b) = p_{12}(\phi)$ holds, where $\phi = |b - a|$ is the angle between the analyzer axes, then (B3) becomes

$$-1 \leq 3p_{12}(\phi) - p_{12}(3\phi) - p_1 - p_2 \leq 0. \quad \text{(B4)}$$

Here, $a$, $a'$, $b$ and $b'$ have been chosen so that

$$|a - b| = |a' - b| = |a' - b'| = \frac{1}{3}|a - b'| = \phi.$$

The upper limits in (B3) and (B4) are experimentally testable without $N$ being known. Inequalities (B3) and (B4) hold perfectly generally for any systems described by OLT. These are new results not previously presented elsewhere. [End of quoted text.]

## 2. CHSH inequality with supplementary assumptions

On page 530, we find derivation of an inequality of similar structure that can be used with real, low-efficiency, detectors. It employs an assumption rather stronger than (B1). This is the "no enhancement" assumption, which can be expressed mathematically in the form:

$$0 \leq p_1(\lambda, a) \leq p_1(\lambda, \infty) \leq 1,$$
$$0 \leq p_2(\lambda, b) \leq p_2(\lambda, \infty) \leq 1, \quad \text{(B5)}$$

where $\infty$ denotes absence of the polariser, $p_1(\lambda, \infty)$ the probability of a count from detector 1 when the polariser is absent and the emission is in state $\lambda$, and $p_2(\lambda, \infty)$ likewise for detector 2.

Using the same arguments as before, we find (B3) replaced by:

$$-p_{12}(\infty,\infty) \leq p_{12}(a,b) - p_{12}(a,b') + p_{12}(a',b) + p_{12}(a',b')$$
$$- p_{12}(a', \infty) - p_{12}(\infty, b) \leq 0. \quad \text{(B6)}$$

and (B4) (the version for use when we have rotational invariance) replaced by:

$$-p_{12}(\infty,\infty) \leq 3p_{12}(\phi) - p_{12}(3\phi)$$
$$- p_{12}(a', \infty) - p_{12}(\infty, b) \leq 0. \quad \text{(B7)}$$

When the two sides of the experiment are symmetrical, after dividing through by $p_{12}(\infty, \infty)$ the right-hand inequality leads to the $S_C$ test of table II above.

## 3. Theorem from Clauser and Horne's original Appendix A

The above derivations depend on the following theorem, proved on page 530 of Clauser and Horne's paper:

Given six numbers $x_1, x_2, y_1, y_2, X$ and $Y$ such that

$$0 \leq x_1 \leq X,$$
$$0 \leq x_2 \leq X,$$
$$0 \leq y_1 \leq Y,$$
$$0 \leq y_2 \leq Y,$$

then the function $U = x_1 y_1 - x_1 y_2 + x_2 y_1 + x_2 y_2 - Y x_2 - X y_1$ is constrained by the inequalities

$$-XY \leq U \leq 0.$$

## APPENDIX C: INTEGRATION OF THE STANDARD REALIST FORMULA

We require to integrate equation (5),

$$P_{AB,R}(a,b) = \int_0^\pi \frac{d\lambda}{\pi} \cos^2(\lambda - a) \cos^2(\lambda - b),$$

using elementary maths.

Now the trigonometric identity

$$\cos(A+B) + \cos(A-B) \equiv 2\cos A \cos B$$

tells us that

$$P = \int_0^\pi \frac{d\lambda}{4\pi}\{\cos((\lambda-a)+(\lambda-b)) + \cos((\lambda-a)-(\lambda-b))\}^2$$
$$= \int_0^\pi \frac{d\lambda}{4\pi}\{\cos^2(2\lambda - a - b))$$
$$+ 2\cos(2\lambda - a - b).\cos(b-a) + \cos^2(b-a)\}.$$

The identity $\cos^2 A \equiv \frac{1}{2}(1 + \cos 2A)$ then gives us:

$$P = \int_0^\pi \frac{d\lambda}{4\pi}\{\frac{1}{2} + \frac{1}{2}\cos(4\lambda - 2a - 2b)$$
$$+ 2\cos(2\lambda - a - b).\cos(b-a) + \cos^2(b-a)\}.$$

The second and third terms contribute zero to the integral, being cosines integrated over complete periods. We are left with:

$$P = \frac{1}{4\pi}(\frac{\pi}{2} + \pi\cos^2(b-a))$$
$$= \frac{1}{8} + \frac{1}{4}\cos^2\phi,$$

which can alternatively be written

$$P = \frac{1}{4} + \frac{1}{8}\cos 2\phi.$$

---

intensity light of these experiments are unable to detect any but the strongest pulses unless there is a certain amount of electromagnetic noise present. Evidence for the importance of noise includes the fact that temperature and applied voltage have a strong effects on the behaviour. (See, for example, a paper by G. Ribordy *et al.*: "Performance of InGaAs/InP avalanche photodiodes as gated-mode photon counters", Applied Optics **37**, 2272-77 (1998).) A useful classical model assumes that all detections occur as a result of the total of the signal and the local noise exceeding some threshold, when integrated over a very short interval (much less than the coherence time of 10 ns or more of Aspect's experiments). One consequence of this model is that a light pulse can be detected at any point while its intensity remains significant, but the probability of detection is, for atomic cascade sources, greatest near the front. It is of interest to note that the process could in theory result in multiple detections. Instrument dead times ensure that in practice these are rare.